# High-resolution Resonance Bragg-scattering spectroscopy of an atomic transition from a population difference grating in a vapor cell

Hai Wang[*], Xudong Yang, Shujing Li, Chunhong Zhang, Changde Xie, Kunchi Peng

*The State Key Laboratory of Quantum Optics and Quantum Optics Devices, Institute of Opto-Electronics, Shanxi University, Taiyuan, 030006, People's Republic of China*

**Abstract**

The laser spectroscopy with a narrow linewidth and high signal to noise ratio (S/N) is very important in the precise measurement of optical frequencies. Here, we present a novel high-resolution backward resonance Bragg-scattering (RBS) spectroscopy from a population difference grating (PDG). The PDG is formed by a standing-wave (SW) pump field in thermal $^{87}$Rb vapor, which periodically modulates the space population distribution of two levels in the $^{87}$Rb D1 line. A probe beam, having the identical frequency and the orthogonal polarization with the SW pump field, is Bragg-scattered by the PDG. Such Bragg-scattered light becomes stronger at an atomic resonance transition, which forms the RBS spectrum with a high S/N and sub-natural linewidth. Using the scheme of the coherent superposition of the individual Rayleigh-scattered light emitted from the atomic dipole oscillators on the PDG, the experimentally observed RBS spectroscopy is theoretically explained.

The techniques of Doppler-free laser spectroscopy [1] such as saturated absorption and two-photon laser spectroscopy invented in the early 1970s have marked impacts on the field of the precision spectroscopy. These techniques have been applied in a variety of modern scientific and technical fields, for example, the stabilization of the laser frequency for cooling atoms [2], the measurements of absolute optical frequency of atoms [3, 4], the test of relativistic time dilation [5], and optical frequency metrology [6]. The accuracy of optical frequency measurements of atomic transitions relies on the full width at half maximum (FWHM) and the signal-to-noise ratio (S/N) of the laser spectrum [7]. However, the measurement precision of the laser spectroscopy using the present saturated absorption technique is limited by some physical and technical factors. Usually, the linewidth limitation of the saturated absorption spectroscopy (SAS) is the atomic natural linewidth [3-4], thus the measurement accuracy of atomic transition frequency will be restricted by the natural linewidth. On the other hand, the sensitivity of SAS is affected by the power of the local-oscillator (probe light). In a paper published by M. Ducloy's group [8] in 1980, it was suggested to find a new laser spectroscopy scheme in which the optical signal of the laser spectrum can be separated from the local oscillator to yield background-free detection, and thus the highest S/N can be realized.

For the past few decades, laser-induced dynamic gratings have been extensively studied owing to its potential applications in scientific investigation and precise measurements. It has been well-known that a strong standing-wave (SW) optical field in an atomic medium can induce various gratings by periodically modulating different physical parameters of atomic systems. The optical lattices loading cold atoms created by the interaction of the off-resonance SW field are the typical atomic density gratings, which can Bragg scatter a light beam [9-11]. In the three-level EIT system with a stronger SW coupling field, the atomic coherence is spatially modulated, thus an electromagnetically induced grating (EIG) [12-14] is formed. The Bragg reflection from the EIG has been experimentally demonstrated in a vapor cell [13-14]. Another type of atomic dynamic gratings induced by SW optical field is the population difference grating (PDG) resulting from the intensity-dependent modulation of atomic population. The idea of the PDG was proposed initially by Horoche and Hartmann [15] in 1972 and then was used to explain the self-cooling phenomenon of two-level atoms [16]. The four-wave-mixing (FWM) effects associated with the PDG were also studied theoretically and experimentally [17-20].

Here, we present a novel high-resolution resonance Bragg-scattering (RBS) spectroscopy from a PDG in a thermal Rb atomic cell. Unlike the Bragg reflection from the EIG via atomic coherence effect in a three-level atomic system [13-14], in our experiment only two levels of Rb atoms are involved, and the frequency of all laser (probe and pump) beams is identical during the frequency scanning, that is why the presented RBS scheme can be used for measuring the atomic spectrum. In contrast to the explanation to the Bragg reflection or diffraction using the modulated absorption and refractive indexes of the atomic medium in the previously presented papers [11-14], in our theoretical model, the Bragg scattered light is regarded as being the coherent superposition of the Rayleigh-scattered optical fields emitted by the individual atomic oscillating dipoles on the PDG. The calculated Bragg-scattering spectrum exhibits a non-Lorentzian function and thus it can reach a sub-natural linewidth when the power broadening induced by optical fields may be neglected. The sub-natural linewidth of ~3.5MHz (FWHM) is obtained in the backward RBS spectrum at low-light-level and the spectrum with a peak power of 3.5 μW and a high S/N of ~2000 is achieved at high-light-level.

The $^{87}$Rb atomic energy levels are shown in Fig.1 (a), |b> and |a> are the ground and excited states, respectively, the transition frequency from |b> to |a> is $\omega_{ba}$. The state |c> is another ground state, a non-radiatively-coupled third state. A horizontally-polarized (x-polarized) SW pump field $\boldsymbol{E}_{sw}(t)=\boldsymbol{E}_{F}(t)+\boldsymbol{E}_{B}(t)$ and a vertically-polarized (y-polarized) probe field $\boldsymbol{E}_{P}(t)=\boldsymbol{e}_{y}E_{P}e^{-i\omega t-ikz}$ both couple to the |b> to |a> transition, where $\boldsymbol{E}_{F}(t)=\boldsymbol{e}_{x}E_{F}e^{-i\omega t-ikz}$ and $\boldsymbol{E}_{B}(t)=\boldsymbol{e}_{x}E_{B}e^{-i\omega t+ikz}$ are x-polarized forward- and backward-propagating pump fields

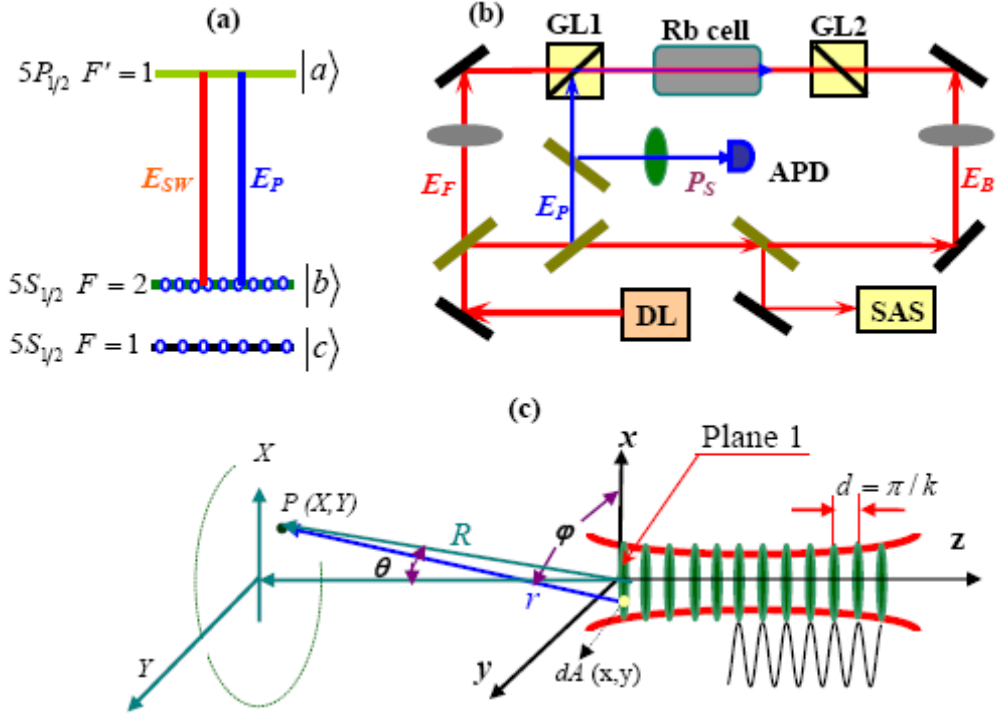

**Fig.1** The essential experimental information. (a) Relevant energy levels of $^{87}$Rb atom. (b) The schematic of experimental set-up. DL is an external-cavity diode laser with a bandwidth ~0.5 MHz during a short scanning time (~50ms). GL1 and GL2 are two Glan-laser prisms with an extinction ratio $10^5$, APD is an avalanche detector, SAS is a rubidium saturated absorption spectrometer. (c) The schematic of RBS spectrum from the PDG, which is formed by a SW pump field.

respectively. $k$ is wave vector, $\mathbf{e}_x$ and $\mathbf{e}_y$ are the unit vectors along x and y axes respectively.

The steady-state population difference $\Delta\rho(z) = \rho_{bb}(z) - \rho_{aa}(z)$ for the atoms with the velocity around $v_z = 0$ (the velocity width $\delta v_z \leq \Gamma/k$) can be written as (see Appendix 1)

$$\Delta\rho(z) = \frac{1}{2} - F - \frac{1}{4}\frac{\Omega_F \Omega_B^* \cos 2kz}{\left(\Delta^2 + (\Gamma/2)^2\right)}(\frac{\Gamma/2}{\gamma'}), \qquad (1)$$

where, $\rho_{bb}(z)$ and $\rho_{aa}(z)$ are the populations in the |b> and |a>, respectively, $F$ is a quantity independent on the coordinate z, $\Gamma$ is the natural linewidth of $^{87}$Rb atoms, $\Omega_F$ and $\Omega_B$ are the Rabi frequencies of fields $E_F(t)$ and $E_B(t)$ respectively, $\Delta = \omega - \omega_{ba}$ is the frequency detuning. The Eq. (1) shows that $\Delta\rho(z)$ is modulated along the direction of the coordinate z

due to the optical pumping of the SW field. At the node positions of $z = (m+1/2)\pi/k$ ($m$=0, 1, 2,...), $\Delta\rho(z)$ reaches the maximum, while at antinode positions of $z = m\pi/k$, $\Delta\rho(z)$ has the minimum, thus the grating period $d$ equals to $\pi/k$ ($\lambda/2$). For the atoms with a speed of $V_z \neq 0$, the periodic modulation of $\Delta\rho(z)$ will not be effectively produced, since the simultaneously resonant interactions of the atoms with $E_F(t)$ and $E_B(t)$ can't occur when the Doppler frequency shifts seen by the two counter-propagating light beams are different.

When the near-resonance probe field interacts with the two-level atoms, the scattered light contains a sum of elastic (Rayleigh) and inelastic components [21, 22]. The inelastic component is incoherent, and doesn't contribute to the Bragg scattering. While, the Rayleigh component resulting from the radiation of the atomic dipole oscillating is coherent, and thus its contribution forms the RBS spectrum in the presented experiment. For the used isotropic two-level atoms driven by the y-polarized probe field $E_P(t)$, the oscillating dipole moment is along the y-direction, and its expectation value can be expressed as (see Appendix 2)

$$\langle \boldsymbol{\mu}_y(t) \rangle = \frac{-|\mu_{ab}|^2 E_P e^{-i\omega t - ikz} \Delta\rho(z) \boldsymbol{e}_y}{3\hbar(\Delta + i\Gamma/2)} + c.c , \qquad (2)$$

where $|\mu_{ab}|$ is the dipole moment for the transition of |a> to |b>. All dipole oscillators in a $x - y$ plane have an identically oscillating frequency $\omega$ and an initial phase $kz$ which are the same with that of the probe field $E_P(t)$ at the position z. The dipole oscillators emit photons into all the directions, which are just the Rayleigh-scattered light. The light field $E_S$ Rayleigh-scattered by one atom can be written as [21-22]

$$E_S = \frac{[\langle \ddot{\boldsymbol{\mu}}_y(t - \frac{r}{c}) \rangle \cdot \boldsymbol{e}_s]\sin\varphi e^{-ikr}}{4\pi\varepsilon_0 c^2 r} \approx \frac{-\omega^2 |\mu_{ab}|^2 E_P e^{-i\omega t - ikz} \Delta\rho(z)}{12\pi\varepsilon_0 c^2 \hbar(\Delta + i\Gamma/2)} \frac{e^{-ikr}(\boldsymbol{e}_y \cdot \boldsymbol{e}_s)\sin\varphi}{r}, \qquad (3)$$

where $\boldsymbol{e}_s$ is the polarization vector of the $E_S$, $r$ is the distance from the emitting point (x, y, z)

to the observed point ($X$, $Y$), $\varphi$ is the angle between $x$ axis and $\vec{r}$ ray [see Fig.1(c)]. For calculating the total scattering field from the collection of atoms, we divide the interaction volume $V$ into many small volume elements with the same volume $\delta V = A\delta z$, where $A$ is the area of the grating plane normal to the $z$ axis and $\delta z$ is the thickness along the $z$ axis. Since the phase of each atomic dipole oscillator ($\langle \boldsymbol{\mu}_y(t) \rangle \propto e^{-i\omega t - ikz}$) only depends on the coordinate $z$, if $\delta z$ is small enough we can consider that all dipole oscillators within $\delta V$ radiate essentially in phase. The Rayleigh-scattered fields from the volume elements $\delta V$ at the node positions are stronger than that from the volume elements placed at antinode positions. The amplitudes of the scattered light vary periodically along the $z$-axis with a period length $d$. If the Bragg condition $2d = \lambda$ is satisfied, the coherently enhanced Rayleigh scattering (Bragg-scattering) will appear in the backward direction. The power $P_S$ of the Bragg-scattered light equals to (see the Appendix 3)

$$P_S = k \frac{P_F P_B P_P}{\left[\Delta^2 + (\Gamma/2)^2\right]^3}, \qquad (4)$$

where, $P_{F(B)}$ and $P_P$ are the powers of the forward (backward) propagating pump fields and the probe field, respectively. From Eq. (4), we can see that the $P_S$ depends on the detuning $\Delta$ and the strongest $P_S$ presents at the atomic resonance, which forms the RBS spectrum. The calculated FWHM linewidth of the RBS spectrum equals to $\Gamma/2$. The physical reason to produce such a narrow spectrum is that the RBS has a non-Lorentzian function $\left[\Delta^2 + (\Gamma/2)^2\right]^{-3}$, which is different from the Lorentzian function $\left[\Delta^2 + (\Gamma/2)^2\right]^{-1}$ with a atomic natural linewidth of $\Gamma$.

The experimental set-up for the RBS spectrum is shown in Fig.1(b). A laser beam, coming from an external-cavity diode laser (LD), is divided into three beams by the beam splitters. One serves as the probe beam, the other two are used for the forward- and backward-propagating pump beams. The Glan prisms placed respectively in the forward and backward pump beams (propagating along $z$ and $-z$ axis) make the polarization of the two optical beams in x-axis direction. The two pump beams overlap in an Rb vapor cell to build a SW (as shown in Fig.1(b)). The probe beam overlaps with the forward pump beam on a polarized-beam-splitter and the output probe beam is polarized along y-direction. The atomic cell is $l=5cm$ long and is wrapped in µ-metal, its resident magnetic field is < 20mGs. The radii ($\rho_0$) of all probe and pump beams are about 1mm at the center of the cell. We scanned the frequency of the LD across the resonance of transition F=2 to F'=1 to measure the backward RBS spectrum of the transition. The FWHM linewidth of the RBS spectrum was calibrated with the half of the frequency splitting (408.3MHz) between the states $5P_{1/2}$ F'=1 and F'=2, which is measured by the saturated absorption spectrometer. Firstly, fixing the forward pump power $P_F$ (300µW) and the probe power $P_P$ (32µW), the RBS spectra are measured under different backward pump powers $P_B$ ($P_B$ =56, 80, 100, 130, 160, 210, 252µW, respectively). The measured results are shown in Fig.2(a)-(g) with the blue curves. From Fig. 2(a) we can see that the FWHM of RBS spectrum is ~3.5 MHz at $P_B$ =56µW. Increasing the power $P_B$, the intensity of the SW pump field is enhanced and the population modulation of the atoms also is strengthen, which must result in stronger RBS signal. Simultaneously, the FWHM linewidth of RBS spectrum gradually increases to ~6MHz due to the influence of the saturated effect. The linewidths of all RBS spectra in Fig.2(a)-(f) are narrower than the natural atomic linewidth of $\Gamma \approx 5.8MHz$. Fig.2(h) is the measured RBS spectrum with $P_F=440µW$, $P_B=300µW$, $P_P=32µW$. In this case the signal intensity of RBS spectrum is increased, while the FWHM linewidth is simultaneously broadened to ~9MHz. When $P_P$ is increased to $P_P = 552µW$ and $P_F$ and $P_B$ are set to $440µW$ and $300µW$ respectively, the signal peak power of RBS is increased to $3.5µW$ and the linewidth is broaden to 12.9MHz [Fig.2(i)]. The S/N in Fig.2(i) is ~2000, which is limited by the dark electronic noise of the detected system. The theoretical prediction shows that if the dark electronic noise is effectively suppressed, the S/N of the RBS spectrum with a high peak power of $3.5\mu W$ can be increased to $\sim 10^6\text{-}10^7$. From Fig.2(a)-(i), it is pointed out that the intensity and the linewidth of the RBS spectrum depend on $P_F$, $P_B$ and $P_P$, In experiments, we can adjust the three powers respectively to obtain the optimal RBS spectrum according to different requirements.

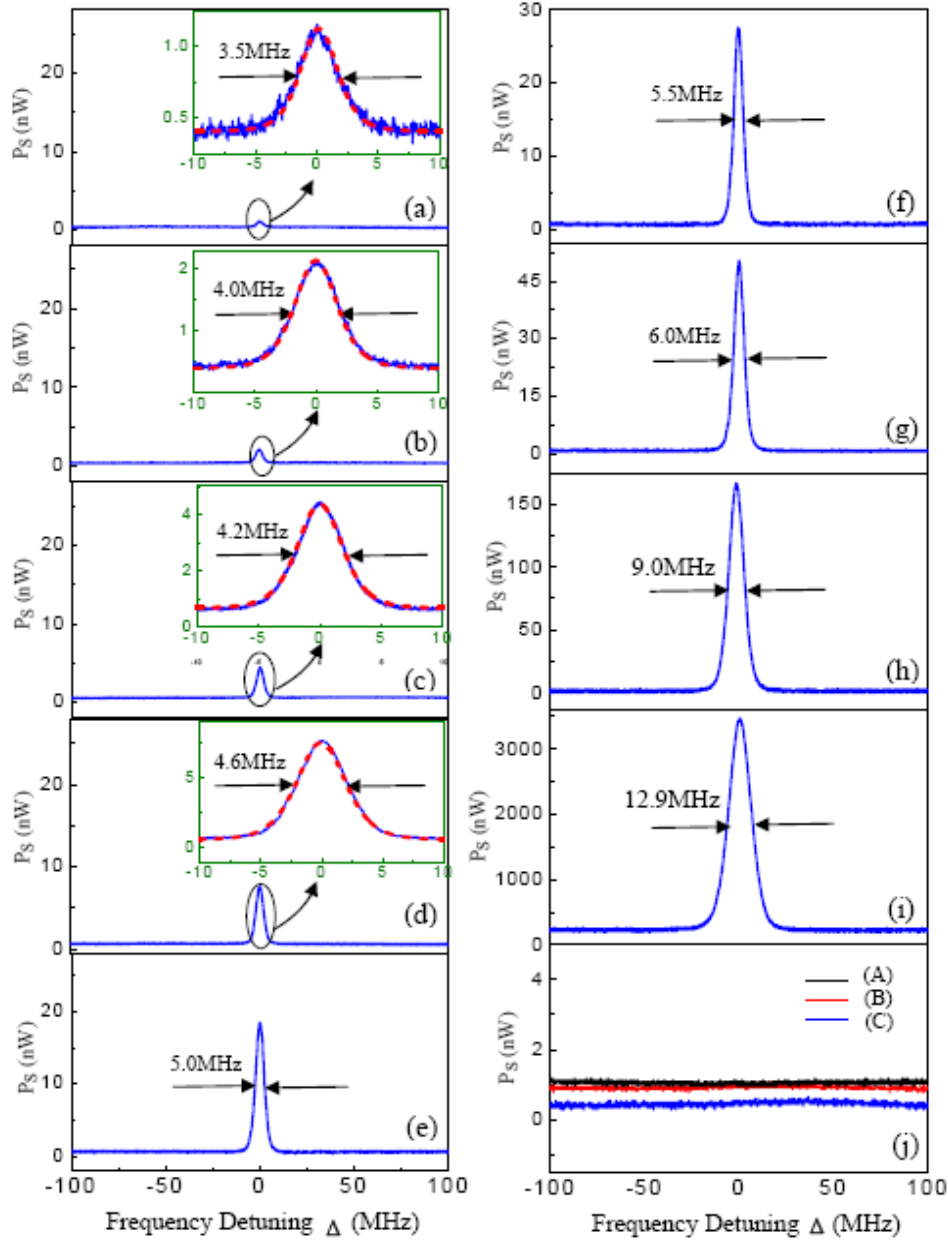

**Fig.2** The measured RBS spectrum for different $P_B$, $P_F$ and $P_P$. (a)-(g), are the measured RBS spectrum at $P_F = 300\ \mu W$ and $P_P = 32\ \mu W$ for $P_B = 56, 80, 100, 130, 160, 210, 252\mu W$, respectively. The red (dotted) curves in insets of Fig.(a)-(d) are the theoretical fits using the function $\left[\Delta^2 + (G\Gamma/2)^2\right]^{-3}$, G is the fitting parameter. (h), is the measured RBS spectrum for $P_F = 440\ \mu W$, $P_P = 32\ \mu W$, and $P_B = 300\mu W$. (i), is the measured RBS spectrum for $P_F = 440\ \mu W$, $P_P = 552\ \mu W$, and $P_B = 300\mu W$. Fig.2 (j) are the measured results when one of three beams [backward and forward pump ($E_B$, $E_F$), and probe $E_P$ beams] is blocked. The powers of the three beams are set at $P_F = 300\ \mu W$, $P_B = 100\ \mu W$, $P_P = 100\ \mu W$. The red, black and blue traces correspond to the cases without $E_B$, $E_F$, or $E_P$, respectively.

It is interesting that when any one of the three beams ($E_F(t)$, $E_B(t)$ and $E_P(t)$) is blocked, the RBS spectrum will totally disappear. The black (red) curve in Fig.2(j) is the measured RBS signals in the absence of $E_F(t)$ ($E_B(t)$) beam, where the RBS spectroscopy signal disappears, that is because the PDG no longer exists without the SW pump field built by both $E_F(t)$ and $E_B(t)$ beams. The blue curve in Fig.2(j) is the measured Bragg-scattered light in the absence of $E_P(t)$ beam, of course, the RBS spectrum also doesn't exist because there is no incident probe field.

Fig.3(a) is the measured peak powers $P_{SP}$ ($P_S|_{\Delta=0}$) of RBS spectra as the function of the $P_B$, the inset shows its linewidth versus $P_B$. The peak powers $P_{SP}$ go slowly up when the $P_B$ increases for $P_B < 160\mu W$, which is not in agreement with the theoretical prediction from Eq.(4). We consider that perhaps the effect of the atomic absorption to the RBS light, which is not involved in Eq.(4), can not be neglected for such a weak probe beam. When $P_B > 160\mu W$, the peak powers $P_{SP}$ increase approximately proportional to $P_B$ linearly, which is in agreement with Eq.(4). Using the data on the linearity fit of Fig.3 (a), we calculated the ratio value $k = \frac{P_{SP}(\Gamma/2)^6}{P_F P_B P_P} \approx 1.83 \times 10^{43} Hz^6/W^2$. Fig.3 (b) and (c) are the measured peak powers $P_{SP}$ of RBS spectroscopy as the function of the $P_F$ and $P_P$, respectively. The insets are the corresponding linewidths. The data in Fig.3(b) (Fig.3(c)) show that $P_{SP}$ linearly depend on $P_F$ ($P_P$), which is in good agreement with the theoretical prediction from Eq.(4). We calculated

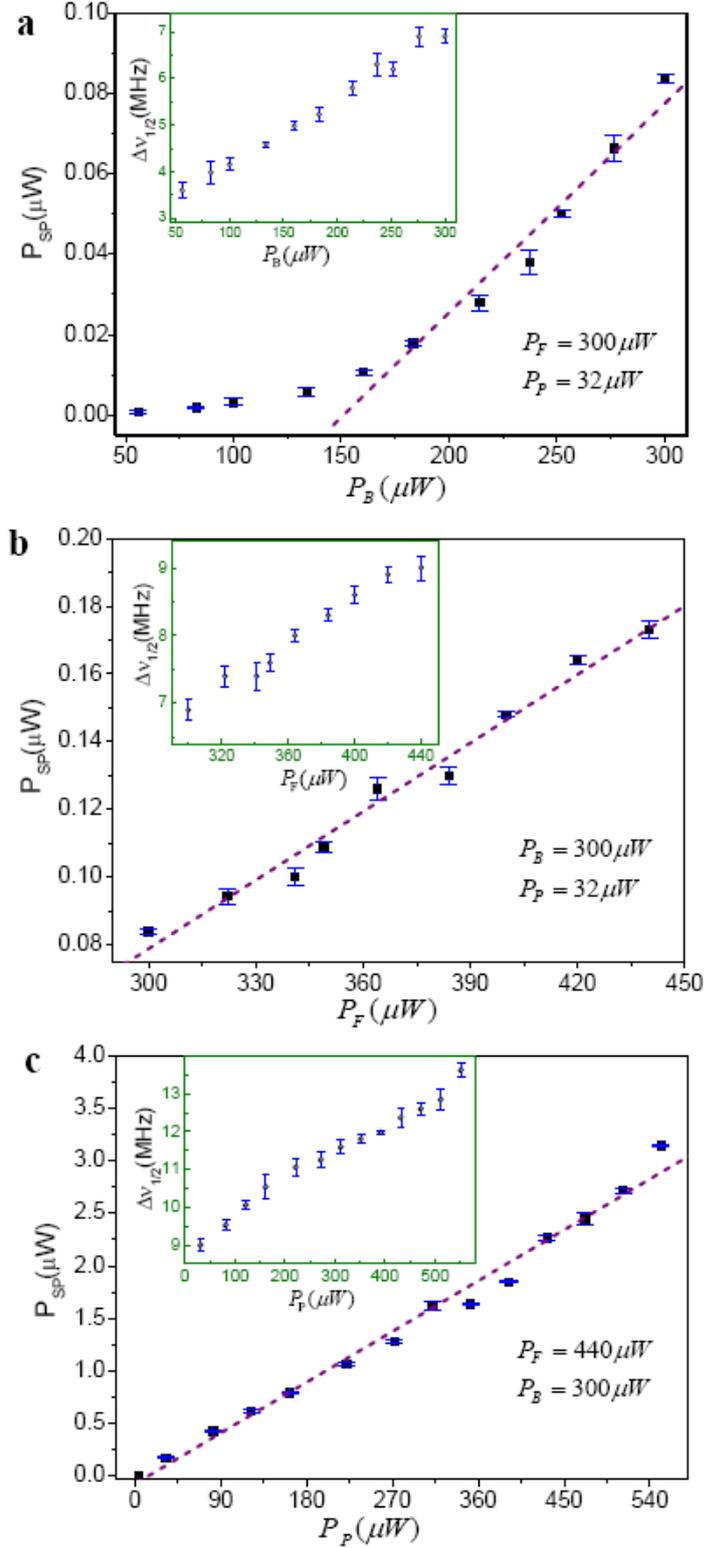

**Fig.3** The dependences of the peak power $P_{SP}$ (i.e. $P_S|_{\Delta=0}$) of the RBS spectrum on $P_B$ (a), $P_F$ (b) and $P_P$ (c), respectively. The insets in a, b and c show the dependences of the linewidth of the RBS spectroscopy on $P_B$, $P_F$ and $P_P$, respectively.

$k \approx 2.8 \times 10^{43} Hz^6/W^2$ in Fig.3(b) and $k \approx 2.8 \times 10^{43} Hz^6/W^2$ in Fig.3(c). The $k$ values

calculated from Fig.3(a), Fig.3(b) and Fig.3(c) have the same order of magnitude, which demonstrate that the given simple theoretical model for explaining the experimental results is reasonable. From Eq.[4] with parameters $N_0 \approx 2\times10^{12}/cm^3$ (corresponding to the cell temperature of ∼91℃ ), the theoretically calculated result is $k \approx 1.75\times10^{45} Hz^6/W^2$, which is larger than that obtained from the experiment. The discrepancy between the theoretical and the experimental values are mainly due to the following reasons: (1) In our simple theoretical model, the atomic collisions are neglected. Actually, the atomic collisions may lead some of the atoms with near zero velocity to escape off such a small region $\delta V$, which will induce a loss of the scattered light intensity. (2) The absorptions of atoms for the scattered light are also neglected. A more detailed theoretical calculation is going on.

For proving whether the varying of $P_F$ ($P_B$) may affect on the centre of the RBS spectroscopy, we build another set of the experimental set-up having the identical configuration with the old one. In Fig.4 (A), (B) and (C), the curves c (red) are the RBS spectra measured with the old set-up under different $P_F$ and $P_B$ ($P_P$=56μW). The curves b (blue) in Fig.4 (A), (B) and (C) are the RBS spectra measured with the new one at a set of fixed powers of $P_F$=300μW, $P_B$=56μW and $P_P$=32μW. It is shown that the centers of the RBS spectra are not shifted almost when $P_F$ and $P_B$ are changed (see red curves), and always overlap with the blue curves measured by an other set-up for a set of fixed powers. It means that the central position and the line shape of the RBS spectra don't depend on either the powers of the optical fields. The feature is specially useful if we would like to develop a RBS spectroscopy based on the atomic PDG.

We would like to point out that the satisfied condition in the presented RBS scheme is different from the phase matching condition in DFWM [17-20](see Appendix 4 for details). The achieved S/N of the RBS spectra in the experiment are limited by the background electronic noise of the APD detector and the measurement interval (the interval of scanning 15MHz is about 1ms). By reducing the background electronic noise of the APD detector and improving the measurement method (for example, using the locking technique in amplifier), even higher S/N may be achieved. Unlike the saturated absorption spectrum, the RBS spectrum can be separated from the local oscillator (probe beam), and then it can immunize the affects of local oscillator power to yield background-free, signal-shot-noise-limited detection [8]. The properties of the presented RBS spectroscopy provide us a potential to develop a high resolution spectroscopy and a new tool for accurately measuring the atomic or molecular resonance transitions. In addition, the RBS technique also can be extended to detect a transition of closed-two-level atoms (see Appendix 3 for details). Up to now, the measurement for ultra-narrow (dipole- or spin-forbidden) transitions of closed-two-level atoms is still a technical challenge even if its frequency is already known [24], that is because the interaction of the light with atoms for a forbidden transition is too weak. By increasing the power of $P_F$ ($P_B$, or $P_P$), it is possible to achieve a sensitive RBS spectroscopy of the ultra-narrow forbidden resonance based on the presented system. We believe that the new scheme of PDGRBS has an impact on the precise frequency measurement and can provide a convenient frequency standard for the narrow line cooling of atoms.

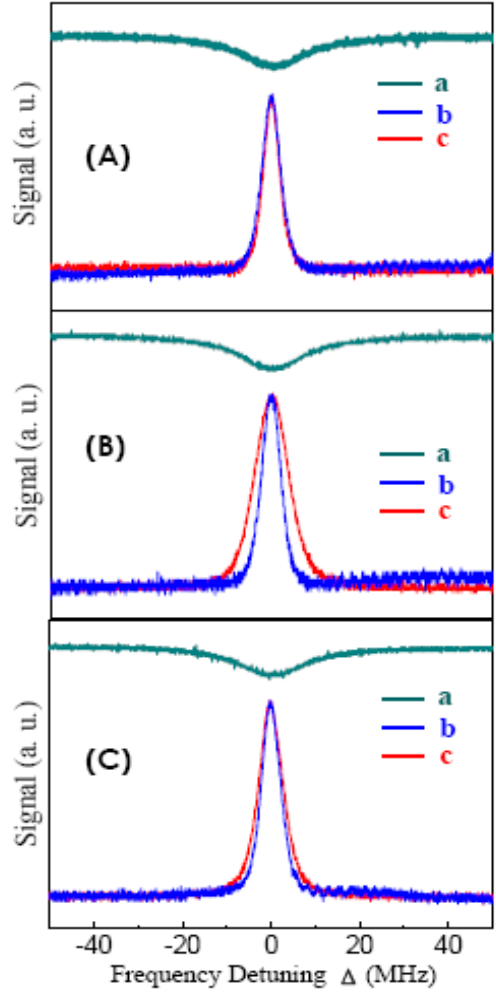

**Fig.4** The comparison of the RBS spectra measured with the old experimental set-up at different $P_F/P_B$ with that measured from the new one. In (A), (B) and (C), curves a (green) are the saturated absorption spectra, curves c (red) are the measured RBS spectrum from the old experimental set-up. The corresponding powers of the light fields are $P_F$=300μW, $P_B$=80μW, and $P_P$=56μW for (A), $P_F$=300μW, $P_B$=300μW, and $P_P$=56μW for (B), $P_F$=150μW, $P_B$=300μW, and $P_P$=56μW for (C). The curves b (blue) are the RBS spectrum measured with the new experimental set-up at Pp=32μW, $P_B$=56μW and $P_F$=300μW.

**Appendix:**

**1. The calculation for the population difference $\Delta\rho(z)$**

The interaction Hamiltonian of the two-level atoms coupled to the x-polarized SW pump field $E_{sw}(t)$ and a y-polarized probe field $E_P(t)$ can be written as

$$\hat{V} = -\mathbf{\mu}_{ab} \bullet [E_{sw}(t) + E_P(t)] |a\rangle\langle b| + \text{c.c.},$$

where $E_{sw}(t) = E_F(t) + E_B(t)$, $E_F(t) = e_x E_F e^{-i\omega t - ikz}$ and $E_B(t) = e_x E_B e^{-i\omega t + ikz}$ are x-polarized forward- and backward-propagating pump fields, respectively, while, $E_P(t) = e_y E_P e^{-i\omega t - ikz}$ is y-polarized forward-propagating probe field, $\mathbf{\mu}_{ab} = \langle a|-e\hat{r}|b\rangle$ is the dipole moment. When the three optical fields $E_F(t)$, $E_B(t)$ and $E_P(t)$ interact with moving atoms, the frequency detunings of $E_F(t)$ and $E_B(t)$ are different due to the Doppler frequency shifts. Only for these atoms with the velocity around $v_z = 0$ (velocity width $\delta v_z \leq \Gamma/k$), the Doppler frequency shifts can be regarded as zero and may be neglected. In experiment, usually the intensities of the pump fields $E_F$ and $E_B$ are much stronger than that of the probe field $E_P$, thus we can neglect the influence of $E_P$ in the calculation of the $\Delta\rho(z)$ induced by optical fields. Using the density matrix equations for the two-level atom with a non-radiatively-coupled level given in Ref.[23], we calculated the steady-state $\Delta\rho(z)$ of the atoms with a velocity around $v_z = 0$, which is approximately expressed by

$$\Delta\rho(z) = \frac{1}{2} - F - \frac{1}{4}\frac{\Omega_F \Omega_B^* \cos 2kz}{\left(\Delta^2 + (\Gamma/2)^2\right)}\left(\frac{\Gamma/2}{\gamma'}\right),$$

where $F = \frac{(|\Omega_F|^2 + |\Omega_B|^2)}{8(\Delta^2 + (\Gamma/2)^2)}\left(\frac{\Gamma/2}{\gamma'}\right)$ is a quantity independent of the coordinate z, $\Omega_F = E_F \mu_{ab}^x/\hbar$ ($\Omega_B = E_B \mu_{ab}^x/\hbar$) is the Rabi frequency of $E_F(t)$ ($E_B(t)$), $\mu_{ab}^x$ is the matrix element of dipole moment, $\left(\mu_{ab}^x\right)^2 = \left(\mu_{ab}^y\right)^2 = \left(\mu_{ab}^z\right)^2 = |\mu_{ab}|^2/3$ [22, 23], $\Delta = \omega - \omega_{ba}$ is the frequency detuning, $\Gamma$ is the decay rate of excited state |a>, i.e. the natural linewidth of $^{87}$Rb D1 line, $\gamma'$ is the dephasing rate between |b> and |c>.

## 2. The calculation for the oscillating dipole moment

The expectation value of the oscillating dipole moment equals to [23]:

$$\langle\boldsymbol{\mu}(t)\rangle = \frac{-\boldsymbol{\mu}_{ab}(\boldsymbol{\mu}_{ba}\bullet[\boldsymbol{E}_P(t)+\boldsymbol{E}_{sw}(t)])\Delta\rho(z)}{\hbar(\Delta+i\Gamma/2)} + c.c$$

$$= \frac{-\boldsymbol{\mu}_{ab}\boldsymbol{\mu}_{ba}[E_P e^{-i\omega t-ikz}\boldsymbol{e}_y + \left(E_F e^{-i\omega t-ikz} + E_B e^{-i\omega t+ikz}\right)\boldsymbol{e}_x]\Delta\rho(z)}{\hbar(\Delta+i\Gamma/2)} + c.c,$$

Since the presented two-level atomic system is isotropic, the direction of the atomic polarization $\langle\boldsymbol{\mu}(t)\rangle$ is parallel with that of the total applied field $\boldsymbol{E}(t) = \boldsymbol{E}_P(t) + \boldsymbol{E}_{sw}(t)$ [23], i.e. $\langle\boldsymbol{\mu}(t)\rangle = \alpha\boldsymbol{E}(t)$, $\alpha$ is the atomic polarizability. In this case, all off-diagonal elements of the $\boldsymbol{\mu}_{ab}\boldsymbol{\mu}_{ba}$ tensor are zero, and the diagonal elements equal to $|\mu_{ba}|^2/3$. So, the y-direction expectation value of the oscillating dipole moment is given by

$$\langle\boldsymbol{\mu}_y(t)\rangle = -\frac{|\mu_{ab}|^2 E_P(t)\Delta\rho(z)\boldsymbol{e}_y}{3\hbar(\Delta+i\Gamma/2)} + c.c,$$

The above equation means that the y-polarized dipole oscillating will vanish when the y-polarized probe beam is blocked, which is in good agreement with the experimental result in Fig.2(j).

## 3. The calculation for the total power of the backward light scattered from a collection of atoms.

According to the Eq.(3), the Rayleigh-scattered field $\boldsymbol{E_s}$ arriving at the point $P(X, Y)$ from the volume element $\delta V = A \delta z$ at $z=0$ [plane 1 in Fig.1(c)] can be written as:

$$E_{S1} = \iint_A \frac{-\omega^2 |\mu_{ab}|^2 E_p e^{-i\omega t} \Delta\rho(z)}{12\pi\varepsilon_0 c^2 \hbar(\Delta + i\Gamma/2)} \frac{e^{-ikr}[\boldsymbol{e}_y \cdot \boldsymbol{e}_s]\sin\varphi}{r} N\delta z dA \ ,$$

where, $N$ is the atomic number density, $r$ is the distance from a small scattering area $dA$ at a position $(x, y, z=0)$ to a point $P(X,Y)$ on the observation screen. Using the Fraunhofer approximation and calculating the integral over the cross section $A = \pi\rho_0^2$, the expression of the y-polarized ($\boldsymbol{e}_y \cdot \boldsymbol{e}_s \approx 1$) backward ($\sin\varphi \approx 1$) scattering field is obtained

$$E_{S1} = \frac{2Ae^{-i\omega t - ikR}}{R} \frac{J_1(k\rho_0 \sin\theta)}{k\rho_0 \sin\theta} \frac{-\omega^2 |\mu_{ab}|^2 E_P}{12\pi\varepsilon_0 c^2 \hbar(\Delta + i\Gamma/2)} \Delta\rho(z) N\delta z \ ,$$

$\theta = \arcsin\left(\sqrt{X^2 + Y^2}/R\right)$ (See Fig.1(c)), $R$ is the distance from the central point $(x=0, y=0, z=0)$ on plane 1 [Fig.1(c)] to the point $P(X,Y)$, $J_1$ is the first order Bessel functions. In the presented experiment, the radius of the laser beams $\rho_0$ is large enough (*1mm*) and thus $\theta$ is very small. So the scattered light from a volume element $\delta V = A \delta z$ can be considered as a plane wave along $-z$ axis. For the grating volume element placed at position $z_i$, the initial phase of the oscillating dipole moments should be $-ikz_i$, the phase of the radiation arriving the position of $z=0$ is $-i2kz_i$. The total scattered field $E_S$ equals to the sum of the field $E_{Si}$ scattered from respective volume elements:

$$E_S = \sum E_{Si} = H(\theta, R) \frac{-\omega^2 |\mu_{ab}|^2 E_p}{12\pi\varepsilon_0 c^2 \hbar(\Delta + i\Gamma/2)} \int_0^l N\Delta\rho(z) e^{-i2kz} dz \ , \qquad (5)$$

where $H(\theta, R) = [2Ae^{-i\omega t-ikR} J_1(k\rho_0 \sin\theta)]/(R k\rho_0 \sin\theta)$, $l$ is the length of the Rb cell. Actually, the atoms will incessantly escape out and come into the region due to the thermal motion. Thus, only the atoms with the velocity $v_z \leq \frac{\lambda}{10}/\tau$ are located within a range of $\Delta z \leq \lambda/10$ when they go through an in-phase region with a transient time $\tau = 2\rho_0/\bar{u}$. The in-phase region means that the phase of each dipole oscillators in the region are essentially identical. $\bar{u}$ is the average atomic velocity along *x-y* plane. These atoms can be regarded as being equally confined in the in-phase region $\Delta V \leq A\lambda/10$. Substituting $N = \int_{-\lambda/10\tau}^{\lambda/10\tau} N(v_z) dv_z$ into Eq. (5), we have

$$E_S = H(\theta, R)\left[\frac{-\omega^2|\mu_{ab}|^2 E_P}{12\pi\varepsilon_0 c^2\hbar(\Delta + i\Gamma/2)}\int_0^l \Delta\rho(z)e^{-i2kz}dz \int_{-\lambda/10\tau}^{\lambda/10\tau} N(v_z)dv_z\right]$$

Substituting the $\Delta\rho(z)$ of Eq.(1) into the above expression and integrating over the area of the observation screen, we obtain the total power of the scattered light:

$$P_S = \frac{1}{2}c\varepsilon_0\int(2|E_S|)^2 dS = \frac{\left[\int_{-\lambda/10\tau}^{\lambda/10\tau} N(v_z)dv_z\right]^2 l^2 A\omega^2|\mu_{ab}|^2|\Omega_P|^2|\Omega_F|^2|\Omega_B|^2(\Gamma/2)^2}{3\cdot 2^7 c\varepsilon_0\left[\Delta^2 + (\Gamma/2)^2\right]^3 \gamma'^2} = k\frac{P_F P_B P_P}{\left[\Delta^2 + (\Gamma/2)^2\right]^3},$$

where $P_{F(B,P)}$ is the power of $E_F$ ($E_B$, $E_P$) field, $k = \frac{\left(\int_{-\lambda/10\tau}^{\lambda/10\tau} N(v_z)dv_z\right)^2 l^2\omega^2|\mu_{ab}|^8(\Gamma/2)^2}{6^4(c\varepsilon_0)^4\hbar^6 A^2 \gamma'^2}$.

Similarly, the expression for the case of a closed two-level atom is calculated:

$$P_S = k\frac{P_F P_B P_P}{\left[\Delta^2 + (\Gamma/2)^2\right]^3}, \text{ where, } k = \frac{\left(\int_{-\lambda/10\tau}^{\lambda/10\tau} N(v_z)dv_z\right)^2 l^2\omega^2|\mu_{ab}|^8}{3^4(c\varepsilon_0)^4\hbar^6 A^2}.$$

**4. The difference between the Bragg condition and the phase matching condition of DFWM.**

According to the solid-state physics, the Bragg condition is

$$\vec{k}_S = \vec{k}_P + (\vec{k}_B - \vec{k}_F)$$

where $\vec{k}_F$, $\vec{k}_B$, $\vec{k}_P$ and $\vec{k}_S$ are the wave vectors of $E_F$, $E_B$ $E_P$ and Bragg-scattered field $E_S$. In our experiment configuration we have $k_F = -k_B = k_P = -k_S$, so the Bragg condition is satisfied. In the general FWM scheme [17-20], the phase matching condition requires $\vec{k}_F + \vec{k}_B - \vec{k}_P - \vec{k}_S = 0$, which is not consistent with the Bragg condition. Further more, we try to understand whether the Bragg scattering in the presented scheme may be interpreted to be a degenerate four-wave mixing (DFWM) process. From the Eq.(5) and Eq.(1), we can derive that $E_S$ of the backward Bragg field is proportional to $E_F E_B^* E_P$, and the phase matching condition corresponding to DFWM should be $\vec{k}_S = \vec{k}_F - \vec{k}_B + \vec{k}_P$, which will yield $k_S = 3k_F$ according to our experiment configuration ($k_F = -k_B = k_P = -k_S$). Such requirement can't be satisfied in our experimental scheme. So the Bragg scattering in the presented work is not a DFWM process.

**Acknowledgement:** We acknowledge funding support from the National Natural Science Foundation of China (No. 60325414, 10874106, 60736040, 60821004), and the 973 Program (2006CB921103). Corresponding author H. Wang´s e-mail address is wanghai@sxu.edu.cn.